# Isotopic effect of proton conductivity in barium-zirconates for various hydrogen-containing atmospheres

M. Khalid Hossain[1, 2, *], T. Yamamoto[1], K. Hashizume[1]

[1]*Department of Advanced Energy Engineering Science, Interdisciplinary Graduate School of Engineering Science, Kyushu University, Fukuoka 816-8580, Japan.*
[2]*Atomic Energy Research Establishment, Bangladesh Atomic Energy Commission, Dhaka 1349, Bangladesh*

*Correspondence: khalid.baec@gmail.com, khalid@kyudai.jp

**Abstract**

Among various perovskite proton conducting oxides, Y-doped $BaZrO_3$ perovskite is a promising material for electrochemical hydrogen devices due to the good chemical stability and higher proton conductivity at higher operating temperatures like 500-800 °C. For the practical application of the functional $BaZrO_3$ proton conductor in the electrochemical hydrogen devices, its necessary to understand the isotopic effect of proton conductivity. To understand the isotopic effect of proton conductivity in the barium zirconates, in this study, the proton conductivity in the Ar, (Ar + 4% $H_2$), (Ar + 4% $D_2$), (Ar + $H_2O$), (Ar + $D_2O$), and $O_2$ atmospheres were measured for two different compositions: $BaZr_{0.9}Y_{0.1}O_{2.95}$ (BZY), and $BaZr_{0.955}Y_{0.03}Co_{0.015}O_{2.97}$ (BZYC) in the temperature range from 500 °C to 1000 °C. By comparing the obtained results, a significant difference in sinterability, conductivity, and the isotopic effect was observed due to the co-doping of the Co element in the $BaZr_{1-x}Y_xO_{3-\alpha}$ proton conductor.

## 1 Introduction

Power production using hydrogen is one of the sustainable and safe energy generation methods [1,2]. Hydrogen energy, as well as fusion energy, is highly expected as a future energy source to solve the roots of environmental and fuel depletion problems [2–7]. Hydrogen isotopes such as deuterium (D) and tritium (T) are used as fuel for fusion power generation. The fusion reactions that are usable in fusion reactors are given below (Eqs. (1)-(4)):

$$D + T \rightarrow {}^4He + n + 17.6 \text{ MeV} \tag{1}$$

$$D + {}^3He \rightarrow {}^4He + n + 18.3 \text{ MeV} \tag{2}$$

$$D + D \rightarrow T + p + 3.0 \text{ MeV} \tag{3}$$

$$D + D \rightarrow {}^3He + n + 3.3 \text{ MeV} \tag{4}$$

Among the above fusion reactions, the DT reaction shown in Eq. (1) can be obtained with the lowest input energy and the generated output energy is large. Therefore, the current research and development of fusion is the DT fusion reactor [4].

Oxide ceramic materials have excellent chemical and physical features such as hardness, incombustibility, and rustproofness when compared with metals and plastic materials [8–12]. Oxide ceramic materials can be exposed under extremely severe conditions such as high temperature, corrosive, and attritional environments. Most common ceramic materials including oxide ceramics resist these severe environments and have better performance than metals and plastic materials. Therefore, the oxide ceramic materials can be widely applied in the area of the new energy system, information system, and recent communication devices [9], also could be used under extreme conditions like nuclear fusion reactor environment [13]. Proton conductive oxide ceramics material could be used as functional electrochemical materials for the nuclear reactor DT fuel circulation system, scientists are currently conducting in-depth research on various properties of proton conductive oxides in this regard [14–16].

In our previous studies, we reported the hydrogen solubilities and diffusivities behavior along with the distribution of hydrogen in the outer and inner portion of the BZY and BZYC proton conducting oxide materials using the tritium imaging plate (TIP) technique (**Fig. 1**) [5,6,17]. But for the practical application of proton conducting oxides in the fusion reactor's fuel circulation system, it is necessary to understand the hydrogen conductivity behavior along with the isotopic effect, but data are not sufficient in the literature. For this reason, in this study, we report the conductivity behaviors of BZY and BZYC in various hydrogen-containing atmospheres especially the deuterium and heavy water atmospheres.

## 2 Proton dissolution and conduction mechanism in BaZrO₃

$BaZrO_3$ itself does not show high proton conductivity but has high proton conductivity by substituting the zirconium ion ($Zr^{4+}$) at the B site with a trivalent cation, i.e., $In^{3+}$, $Yb^{3+}$, $Y^{3+}$, $Gd^{3+}$, etc. Among them, it has been reported that when doped with yttrium ion ($Y^{3+}$), it exhibits the highest proton conductivity [18–20]. Regarding the doping amount, it has been reported that in $BaZr_{1-x}Y_xO_{3-\alpha}$, the proton conductivity increases as the doping amount of Y increases up to x = 0.2, i.e., 20% [21,22]. In this case, $Zr^{4+}$ replaces by two $Y^{3+}$ and one oxygen vacancy ($V_O^{\bullet\bullet}$) is introduced. This oxygen vacancy creation could be described as a point defect using the Kröger-Vink notation equation, given by Eq. 5 [23].

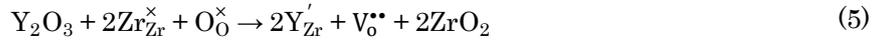

$$Y_2O_3 + 2Zr_{Zr}^{\times} + O_O^{\times} \rightarrow 2Y_{Zr}' + V_O^{\bullet\bullet} + 2ZrO_2 \qquad (5)$$

In other words, Y-doped $BaZrO_3$ has a perovskite structure with oxygen vacancies as shown in **Fig. 2**, and the oxygen vacancies contribute to proton conduction. The $BaZrO_3$ material is a mixed conductor that conducts oxide ions in addition to protons, and there are also reports of electron conduction. It is thought that proton conduction is dominant in the low-temperature range and oxide ions are dominant in the high-frequency range.

Perovskite proton conductive oxides generally take protons not only from the hydrogen atmosphere but also from the water vapor atmosphere and exhibit proton conductivity. In the case of acceptor-doped perovskite-structured oxides, the uptake of protons from hydrogen (Eq. (6)) and water vapor (Eq. (7)) is expressed by the following Kröger-Vink notation equations [24,25]:

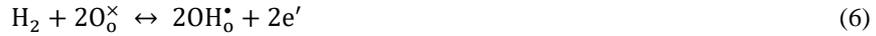

$$H_2 + 2O_O^{\times} \leftrightarrow 2OH_O^{\bullet} + 2e' \qquad (6)$$

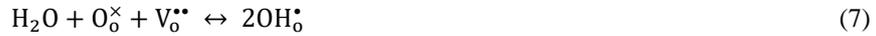

$$H_2O + O_O^{\times} + V_O^{\bullet\bullet} \leftrightarrow 2OH_O^{\bullet} \qquad (7)$$

Here, $O_O^{\times}$ is an oxygen ion with a neutral charge, $OH_O^{\bullet}$ is a hydroxide ion with a single positive charge sitting on an oxygen site, $e'$ is an electron, and $V_O^{\bullet\bullet}$ is an oxygen ion vacancy with two positive charges. Eq. (6) represents that hydrogen is incorporated by binding to oxide ions in the lattice. Eq. (7) represents that protons are taken up through the oxygen vacancies created by doping the acceptor, and the reaction depends on the number of oxygen vacancies [26].

The Vehicle mechanism [27] and the Grothussus mechanism [28–30] are two common mechanisms of proton conduction in the proton conductors. The Vehicle mechanism is the mechanism by which protons move with the diffusion of other large mobile species. As an example, protons in water move to other sites in the form of hydronium ions ($H_3O^+$) using water ($H_2O$) as a vehicle. In perovskite-type oxides, it is thought that $O^{2-}$ moves like a vehicle in the form of $OH^-$. On the other hand, in the Grotthuss mechanism protons move through the structure by transferring protons to oxygen ions at adjacent sites (reorientation) and rotating around oxygen ions by recombination of hydrogen bonds (**Fig. 3**) [31–34].

Since the lattice in the crystal oscillates with a specific frequency and the same applies to the oxygen ions in the perovskite-type oxide, the distance between the oxygen ions in the structure is constantly changing. Protons in perovskite-type oxides are present in the vicinity of oxygen ions in the structure, so they can be treated as $OH_o$. The distance between this $OH_o$ and the oxygen ion at the adjacent site is shortened by the lattice vibration, and the proton can jump to the oxygen ion at the adjacent site. The mechanism by which protons move in this manner is called the Grotthuss mechanism. This mechanism can generally be considered in two steps. One is an operation in which protons transfer to an oxygen ion at an adjacent site, and the other is an operation in which protons rotate around an oxygen ion (reorientation). In the Grotthuss mechanism, by repeating these two steps, protons are conducted through the structure (**Fig. 3**) [34]. A structure with strong hydrogen bonds promotes proton transfer, but it is not advantageous for the reorientation of O-H bonds that require the breaking of hydrogen bonds, and it may reduce the reorientation rate and inhibit long-distance proton transfer [35].

## 3 Experimental

### 3.1 Sintered pellet preparation and characterization

The raw material powders $BaZr_{0.9}Y_{0.1}O_{3-\alpha}$ (BZY) and $BaZr_{0.955}Y_{0.03}Co_{0.015}O_{3-\alpha}$ (BZYC) (prepared by TYK Corporation Ltd., Japan) were used in this study to prepare the pellets by the solid-state reaction method [36]. After weighing about 0.7 g of these powders using an electronic balance (AUY220, manufactured by Shimadzu Corporation), uniaxial compression (load 0.5 t, about 60 MPa, 3 min) was performed for a circular mold sample of diameter 10 mm. Next, put the molded sample in a rubber probe fit, evacuate it with a diaphragm type dry vacuum pump (DAP-6D, manufactured by ULVAC, Inc.), and the cold hydrostatic press (load 8t, about 200MPa, 15 min) was performed. This molded sample was sintered using a molybdenum (Mo) tube furnace at 1640 °C for 20 h in an air atmosphere to get the final sintered pellet.

**Fig. 4** shows the X-ray diffraction patterns for both powder and sintered bodies of the BZY, and BZYC. **Table 1** shows the lattice constants of the powder and sintered body of both samples calculated by refining the diffraction pattern. From the values of the lattice constants of each axis along with corresponding angles obtained by refining each diffraction pattern, it was confirmed that both powder and sintered bodies of BZY and BZYC had a single phase of cubic crystals. The lattice constants (a = b = c) of BZY and BZYC powder were 4.1967 Å, and 4.1939 Å, respectively, whereas the lattice constants (a = b = c) of the BZY and BZYC sintered body were 4.2128 Å and 4.1986 Å, respectively.

**Table 1** also provides the results of density measurement by the Archimedes method for BZY and BZYC samples. The relative density of BZY and BZYC was 98 %, and 99.7%, respectively. As a result of the density measurement, it was confirmed that all samples show very high relative density, and the density of BZYC was improved than BZY.

The SEM images for both powder and sintered samples of BZY, and BZYC are shown in **Fig. 5**. In addition, EDX was performed to identify the constituent elements (**Fig. 6**). In the SEM images (**Fig. 5**) of the powder BZY, the particles are agglomerated with each other and the size of primary grains is not clear but the size of the particles seems to be 0.1 ~1.0 μm as appearing in literature [37–40]. From the outer surface SEM images for all samples, it is observed that the grain boundaries were clear and dense. Also, the rounded particles could be observed with fine particles and grain boundaries. From the fractured surface, it is clear that the surface grains seem to stick together. In addition, from EDX mapping it's clear that all chemical components of the specimen were uniformly distributed throughout the surface (**Fig. 6**).

### 3.2 Fabrication of electrodes

Both sides of the BZY and BZYC sintered pellets were wet-polished with a diamond polishing pad and SiC water-resistant abrasive papers (# 400, # 600, # 800, # 1200, BUEHLER) to achieve a thickness of 0.86 mm. After that, platinum (Pt) paste was applied to both sides of the sample and baked at 1000 °C for 1 h in an air atmosphere. After baking, the side surface of the sample was lightly polished, and it was confirmed that there was no electrical continuity on any sides of the samples [41].

### 3.3 Measurement of conductivity

**Fig. 7** shows a schematic diagram of the conductivity measurement device used in this study. The Pt foil and Pt wire were brought into close contact with the Pt paste-coated surface of the sample. The measurement conditions were an applied voltage of 0.5 or 1.0 V and an operating temperature of 1000 - 500 °C. First, temperature rise to 1000 °C, and then measurement was done up to 500 °C with a 100 °C decrement for each measurement. A temperature controller (Model-DB630, Chino Co. Ltd.) and a digital multimeter (Model-8240, ADC Co. Ltd.) were connected to a personal computer (PC) via general-purpose interface bus (GPIB), and in every minute the measured temperature and current values were recorded automatically in the PC data-storage disk. A DC standard voltage/current generator (Model-TR6142, ADC Co. Ltd.) was used as the power source, and an electrometer (Model-2000, Keithley Instrument) was used for temperature measurement.

The measurement atmospheres were argon (Ar), Ar containing 4% hydrogen isotope (hereinafter, (Ar + 4% $H_2$), and (Ar + 4% $D_2$)), wet Ar containing hydrogen isotope (hereinafter, (Ar + $H_2O$), and (Ar + $D_2O$)), and oxygen ($O_2$) with a flow rate of 100 mL/min. Wet Ar was measured by bubbling $H_2O$ or $D_2O$ with Ar gas. The $H_2O$ and $D_2O$ pressures were considered to be saturated water vapor pressures. The formula for calculating the conductivity is shown below (Eq. 8) [41]:

$$\sigma = \frac{I \times t}{V \times S} \tag{8}$$

Here, σ is the conductivity, $I$ is the steady-state value of the current, $t$ is the thickness of the measured sample, $S$ is the cross-sectional area of the sample, and $V$ is the value of the applied voltage. In addition, BZYC sample was measured three times in each atmosphere to confirm the reproducibility.

## 4 Results

In this paper, the conductivity of BZY in Ar, (Ar + 4% $H_2$), (Ar + 4% $D_2$), (Ar + $H_2O$), (Ar + $D_2O$), and $O_2$ atmospheres are denoted as $\sigma_{BZY\_Ar}$, $\sigma_{BZY\_H_2}$, $\sigma_{BZY\_D_2}$, $\sigma_{BZY\_H_2O}$, $\sigma_{BZY\_D_2O}$, and $\sigma_{BZY\_O_2}$, respectively. Similarly, the conductivity of BZYC in Ar, (Ar + 4% $H_2$), (Ar + 4% $D_2$), (Ar + $H_2O$), (Ar + $D_2O$), and $O_2$ atmospheres are denoted as $\sigma_{BZYC\_Ar}$, $\sigma_{BZYC\_H_2}$, $\sigma_{BZYC\_D_2}$, $\sigma_{BZYC\_H_2O}$, $\sigma_{BZYC\_D_2O}$, and $\sigma_{BZYC\_O_2}$, respectively.

All conductivities data in dry ((Ar + 4% $H_2$) or (Ar + 4% $D_2$)) or wet hydrogen-containing atmospheres ((Ar + $H_2O$) or (Ar + $D_2O$)) for BZY are given in **Table 2,** and **Fig. 8** represent the corresponding Arrhenius plot of conductivities. In the case of BZY, from **Fig. 8(a)**, and **(b)** it is clear that the conductivity in both the dry and wet hydrogen-containing atmospheres showed a higher value than that of the Ar atmosphere. In addition, the conductivity $\sigma_{BZY\_H_2}$ is higher than that of $\sigma_{BZY\_D_2}$ at all measured temperatures (**Fig. 8(a)**), and similarly the conductivity of $\sigma_{BZY\_H_2O}$ is higher than that of $\sigma_{BZY\_D_2O}$ (**Fig. 8(b)**). This is thought to be due to the isotope effect. Furthermore, the conductivity is higher in the $\sigma_{BZY\_O_2}$ atmosphere than in the Ar atmosphere (**Fig. 8(c)**), suggesting conductivity dominated by oxide ions and holes in oxygen containing atmosphere.

All conductivities data in dry ((Ar + 4% $H_2$) or (Ar + 4% $D_2$)) or wet hydrogen-containing atmospheres ((Ar + $H_2O$) or (Ar + $D_2O$)) for BZYC are given in **Table 3,** and **Fig. 9** represent the corresponding Arrhenius plot of conductivities. Similar to BZY, in the case of the BZYC, from **Fig. 9(a)** and **(b)** it is clear that the conductivity in both the dry and wet hydrogen-containing atmospheres showed a higher value than the Ar atmosphere. Besides, the conductivity of $\sigma_{BZYC\_H_2}$ is higher than that of the $\sigma_{BZYC\_D_2}$ at all measured temperatures (**Fig. 9(a)**), and similarly the conductivity of $\sigma_{BZYC\_H_2O}$ is higher than that of $\sigma_{BZYC\_D_2O}$ (**Fig. 9(b)**). This is thought to be due to the isotope effect. Furthermore, the conductivity is higher in the $O_2$ atmosphere than in the Ar atmosphere (**Fig. 9(c)**), suggesting that the conductivity is dominated by oxide ions and holes in oxygen containing atmosphere. In addition, BZYC showed a similar conductivity value as BZY at 900 °C and 1000 °C in both dry and wet hydrogen-containing atmospheres, but it tended to be higher than BZY as the temperature became lower.

The closed blue square data points of **Fig. 10(a)** denote the ratio of the conductivities in dry hydrogen-containing atmospheres for BZY, i.e., between $\sigma_{BZY\_H_2}$ and $\sigma_{BZY\_D_2}$. This ratio is obtained by subtracting the conductivity $\sigma_{BZY\_Ar}$ from the conductivities $\sigma_{BZY\_H_2}$ and $\sigma_{BZY\_D_2}$. Furthermore, the closed black circular data points of **Fig. 10(a)** denote the ratio of the conductivities $\sigma_{BZY\_H_2O}$ and $\sigma_{BZY\_D_2O}$. This ratio is obtained by subtracting the conductivity $\sigma_{BZY\_Ar}$ from the conductivities $\sigma_{BZY\_H_2O}$ and $\sigma_{BZY\_D_2O}$. The ratio of the conductivity of BZY under the dry hydrogen-containing atmosphere ($\sigma_{BZY\_H_2}/\sigma_{BZY\_D_2}$) showed a similar value at 700-1000 °C, and the isotope dependence increased as the temperature decreased. On the other hand, the ratio of conductivity in the wet hydrogen-containing atmosphere for BZY ($\sigma_{BZY\_H_2O}/\sigma_{BZY\_D_2O}$) is in line with the hydrogen isotope ratio (= $\sqrt{2}$) in classical theory [42], confirmed the isotopic effect.

The open blue square data points of **Fig. 10(b)** denote the ratio of the conductivities in wet hydrogen-containing atmospheres for BZYC, i.e., between $\sigma_{BZYC\_H_2}$ and $\sigma_{BZYC\_D_2}$. This ratio is obtained by subtracting the conductivity $\sigma_{BZYC\_Ar}$ from the conductivities $\sigma_{BZYC\_H_2}$ and $\sigma_{BZYC\_D_2}$. Furthermore, the open black circular data points of **Fig. 10(b)** denote the ratio of the conductivities $\sigma_{BZYC\_H_2O}$ and $\sigma_{BZYC\_D_2O}$. This ratio is obtained by subtracting the conductivity $\sigma_{BZYC\_Ar}$ from the conductivities $\sigma_{BZYC\_H_2O}$ and $\sigma_{BZYC\_D_2O}$. It is confirmed that the ratio of the conductivity of BZYC under dry hydrogen-containing atmosphere ($\sigma_{BZYC\_H_2}/\sigma_{BZYC\_D_2}$) tends to increase the isotope dependence as the temperature decreases. In addition, except for the measured value at 500 °C, the ratio of conductivity in the wet hydrogen-containing atmosphere for BZYC ($\sigma_{BZYC\_H_2O}/\sigma_{BZYC\_D_2O}$) is about the same in line with the hydrogen isotope ratio (=$\sqrt{2}$) in classical theory [42].

**Table 4** shows the activation energies of BZY and BZYC under all atmospheres. It is confirmed that the activation energy of BZYC is lower than that of BZY in the reducing atmosphere. On the other hand, in the oxidizing atmosphere, BZY and BZYC showed similar values of activation energies.

## 5 Discussion

For both BZY and BZYC, the conductivity is higher in the dry and wet hydrogen-containing atmospheres than in the Ar atmosphere. In addition, since isotope dependence was confirmed in both dry and wet hydrogen-containing atmospheres, it is considered that BZY and BZYC have proton conductivity, and their proton transport is due to the Grotthuss mechanism [34].

In BZYC, the conductivity at 900 °C and 1000 °C is similar to that of BZY in both dry and wet hydrogen-containing atmospheres, but it is confirmed that the conductivity tended to be higher than that of BZY as the temperature became lower. This is thought to be due to hydrogen solubility. From our previous HT study, we observed that the hydrogen solubility of BZY decreases as the temperature decreases, while that of BZYC increases (**Fig. 1(a)**) [5]. Since BZYC has a higher hydrogen solubility at low temperatures than BZY, it is considered that the conductivity of BZYC became higher than that of BZY as the temperature became lower.

When the isotope dependence of conductivity is treated based on the classical theory with OH bond and OD bond as harmonic oscillators, the difference in the frequency of the bond affects the pre-exponential factor of the diffusion, presented in Eq. (9) [42].

$$\frac{\sigma_{H^+}}{\sigma_{D^+}} \cong \frac{D_{0H^+}}{D_{0D^+}} \cong \left[\frac{m_{D^+}}{m_{H^+}}\right]^{\frac{1}{2}} = \sqrt{2} \qquad (9)$$

Regarding the isotope dependence in the perovskite-type proton conductive oxide, the difference in zero-point vibration energy between the O-H bond and O-D bond oscillators is also considered. Hibino *et al.* [43] measured the conductivity of $CaZr_{0.9}In_{0.1}O_3$ by the complex impedance method and found that the activation energies of conductivity under $Ar + H_2O$ and $Ar + D_2O$ atmospheres differed by the difference in zero-point energy. The difference in the activation energy is thought to be like Eq. (10) [26]. In this case, the isotope dependence of conductivity increases as the temperature decreases.

$$E_D - E_H = \frac{h(\nu_D - \nu_H)}{2} \qquad (10)$$

As a result of this study, both BZY and BZYC tended to become more isotope dependent as the temperature became lower in the dry hydrogen-containing atmosphere, which is consistent with the literature [26,44]. On the other hand, in the wet hydrogen-containing atmosphere the conductivity ratio value is very close to the hydrogen isotope ratio of classical theory ($= \sqrt{2}$) regardless of the measured temperature.

Furthermore, the conductivity is higher in the $O_2$ atmosphere than in the Ar atmosphere, and the difference became larger as the temperature became lower. This may be due to conductivity is dominated by oxide ions and holes, and the reaction formula is shown below (Eq. (11)).

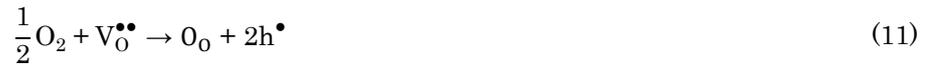

$$\frac{1}{2}O_2 + V_O^{\bullet\bullet} \rightarrow O_O + 2h^\bullet \qquad (11)$$

The conductivity of BZY obtained in this study is higher at 800-1000 °C than the literature values measured by Schober *et al.* in hydrogen atmosphere (sintering conditions: 1715 °C, 30 h, measurement atmosphere: Ar + 4% $H_2$) (**Fig. 8(a)**) [45]. The value was about the same at 600-700 °C and lower at 500 °C. This appears as the difference in the activation energy of conductivity, which is the slope of the Arrhenius plot. The activation energy of conductivity in this study is 1.04 eV, while the literature value is as low as 0.65 eV [45]. The cause of the high activation energy of the sample in this study is considered to be the particle size and the resistance at the electrode interface, but a more detailed investigation is required. Besides, the BZY conductivity obtained in this study is higher at all measurement temperatures than the literature values measured by Slade *et al.* in the wet atmosphere (sintering conditions: 1400 °C, 10 h, measurement atmosphere: $N_2 + H_2O$) (**Fig. 8(b)**) [46]. These differences of current studies conductivities with the literature data might be for the denser sample (due to high-temperature long time sintering), which reduces the intergranular resistance to promote the conductivity.

## 6 Conclusion

In this study, the conductivity of $BaZr_{0.9}Y_{0.1}O_{3-\alpha}$ and $BaZr_{0.955}Y_{0.03}Co_{0.015}O_{3-\alpha}$ was investigated using a 2-terminal DC method to understand the isotopic effect of proton conductivity along with the co-doped sintering aid Co content effect. The BZY and BZYC samples showed higher conductivity both in the dry hydrogen and wet hydrogen atmospheres than in the Ar atmosphere. The conductivity increased even in the $O_2$ atmosphere, and the conductivity difference increased as the temperature decreased. Therefore, it is confirmed that BZY and BZYC are mixed conductors because of the consideration of conductivity by oxide ions and holes along with proton conduction. From the obtained results and discussion, it is clear that sintering aid Co not only improves the sinterability but also improves the conductivity at 500-800 °C. Since isotopic dependence was confirmed both in the dry and wet hydrogen atmospheres, it could be assumed that BZY and BZYC are the potential proton-conducting functional oxide materials for the electrochemical hydrogen devices.

## Acknowledgments

The authors are grateful to the Center of Advanced Instrumental Analysis, Kyushu University, for the support to carry out some parts of this experiment.

## Declaration of interests

The authors declare that they have no known competing financial interests or personal relationships that could have appeared to influence the work reported in this paper.

**Figure captions**

**Fig. 1**. (a) Hydrogen isotope (tritium) IP images of the cross-sections and outer surfaces of the BZY and BZYC materials represents the distribution of hydrogen, (b) Arrhenius plot of the hydrogen solubilities for the BZY, and BZYC, and (c) Arrhenius plot of the hydrogen diffusivities for the BZY, and BZYC [5,6].

**Fig. 2**. Structure of $BaZr_{1-x}Y_xO_{3-\alpha}$.

**Fig. 3**. Proton conduction by Grotthuss mechanism.

**Fig. 4**. XRD patterns for the powder and sintered samples of BZY, and BZYC.

**Fig. 5.** SEM images for powder, sintered outer surface, and sintered inside (fractured surface) of BZY, and BZYC.

**Fig. 6**. EDX mapping for the inside of the BZY, and BZYC sintered samples.

**Fig. 7.** Schematic diagram of the conductivity measurement device.

**Fig. 8**. Conductivity of BZY under: (a) Ar, (Ar + 4% $H_2$), and (Ar + 4% $D_2$) atmospheres, (b) Ar, (Ar + 4% $H_2O$), and (Ar + 4% $D_2O$) atmospheres, and (c) Ar and $O_2$ atmospheres.

**Fig. 9**. Conductivity of BZYC under: (a) Ar, (Ar + 4% $H_2$), and (Ar + 4% $D_2$) atmospheres, (b) Ar, (Ar + 4% $H_2O$), and (Ar + 4% $D_2O$) atmospheres, and (c) Ar and $O_2$ atmospheres.

**Fig. 10.** Conductivity ratio of BZY and BZYC samples.

**Table Captions**

**Table 1.** Some parameters of powders and sintered samples.

**Table 2.** Conductivity of BZY under all atmospheres and temperatures used in this experiment.

**Table 3.** Conductivity of BZYC under all atmospheres and temperatures used in this experiment.

**Table 4.** Activation energy of conductivities under all atmospheres for BZY and BZYC samples

**Table 1.** Some parameters of powders and sintered samples.

| Sample | BZY | BZYC |
| --- | --- | --- |
| Powder structure | Cubic | Cubic |
| Sintered body structure | Cubic | Cubic |
| Powder lattice parameters (Å) | 4.1937 | 4.1939 |
| Sintered body lattice parameters, (a=b=c) (Å) | 4.2128 | 4.1986 |
| Avg. grain size of sintered body (μm) | ~0.7 | ~0.85 |
| Theoretical density (x$10^3$ Kg/m$^3$) | 6.10 | 6.16 |
| Sintered body density (x$10^3$ Kg/m$^3$) | 5.98 | 6.14 |
| % of theoretical density (%TD) | 98.0 | 99.7 |
| Sintered sample open porosity (v%) | 0.1 | 0.1 |
| Sintered sample closed porosity (v%) | 2.1 | 2.5 |

**Table 2.** Conductivity of BZY under all atmospheres and temperatures used in this experiment.

| Operating temp. $T$ (°C) | Conductivity, $\sigma$ (S / m) | | | | | |
| --- | --- | --- | --- | --- | --- | --- |
| | $\sigma_{BZY\_Ar}$ | $\sigma_{BZY\_H_2}$ | $\sigma_{BZY\_D_2}$ | $\sigma_{BZY\_H_2O}$ | $\sigma_{BZY\_D_2O}$ | $\sigma_{BZY\_O_2}$ |
| 500 | 2.76×10$^{-5}$ | 6.59×10$^{-5}$ | 3.78×10$^{-5}$ | 1.34×10$^{-4}$ | 1.03×10$^{-4}$ | 5.90×10$^{-3}$ |
| 600 | 1.10×10$^{-4}$ | 3.29×10$^{-4}$ | 1.95×10$^{-4}$ | 6.42×10$^{-4}$ | 1.90×10$^{-4}$ | 2.76×10$^{-2}$ |
| 700 | 2.34×10$^{-4}$ | 1.23×10$^{-3}$ | 7.45×10$^{-4}$ | 1.23×10$^{-3}$ | 8.49×10$^{-4}$ | 9.54×10$^{-2}$ |
| 800 | 1.18×10$^{-3}$ | 4.18×10$^{-3}$ | 2.46×10$^{-3}$ | 5.18×10$^{-3}$ | 3.47×10$^{-3}$ | 2.29×10$^{-1}$ |
| 900 | 4.36×10$^{-3}$ | 1.18×10$^{-2}$ | 7.57×10$^{-3}$ | 1.42×10$^{-2}$ | 1.08×10$^{-2}$ | 3.92×10$^{-1}$ |
| 1000 | 1.48×10$^{-2}$ | 3.02×10$^{-2}$ | 2.14×10$^{-2}$ | 3.58×10$^{-2}$ | 2.97×10$^{-2}$ | 4.46×10$^{-1}$ |

**Table 3.** Conductivity of BZYC under all atmospheres and temperatures used in this experiment.

| Operating temp. $T$ (°C) | Conductivity, $\sigma$ (S / m) | | | | | |
| --- | --- | --- | --- | --- | --- | --- |
| | $\sigma_{BZYC\_Ar}$ | $\sigma_{BZYC\_H_2}$ | $\sigma_{BZYC\_D_2}$ | $\sigma_{BZYC\_H_2O}$ | $\sigma_{BZYC\_D_2O}$ | $\sigma_{BZYC\_O_2}$ |
| 500 | 1.20×10$^{-4}$ | 2.02×10$^{-4}$ | 1.50×10$^{-4}$ | 3.88×10$^{-4}$ | 2.29×10$^{-4}$ | 7.10×10$^{-3}$ |
| 600 | 3.74×10$^{-4}$ | 9.35×10$^{-4}$ | 5.76×10$^{-4}$ | 1.66×10$^{-3}$ | 1.09×10$^{-3}$ | 3.21×10$^{-2}$ |
| 700 | 1.25×10$^{-3}$ | 2.86×10$^{-3}$ | 1.76×10$^{-3}$ | 4.92×10$^{-3}$ | 3.36×10$^{-3}$ | 1.06×10$^{-1}$ |
| 800 | 3.31×10$^{-3}$ | 6.83×10$^{-3}$ | 4.50×10$^{-3}$ | 1.04×10$^{-2}$ | 8.12×10$^{-3}$ | 2.35×10$^{-1}$ |
| 900 | 7.67×10$^{-3}$ | 1.43×10$^{-2}$ | 1.01×10$^{-2}$ | 2.02×10$^{-2}$ | 1.59×10$^{-2}$ | 4.26×10$^{-1}$ |
| 1000 | 1.64×10$^{-2}$ | 2.82×10$^{-2}$ | 2.12×10$^{-2}$ | 3.55×10$^{-2}$ | 2.70×10$^{-2}$ | 6.32×10$^{-1}$ |

**Table 4.** Activation energy of conductivities under all atmospheres for BZY and BZYC samples

| Operating atmosphere | Activation energy, $E_a$ (eV) | |
|---|---|---|
| | BZY | BZYC |
| Ar | 1.05 | 0.84 |
| Ar+4%$H_2$ | 1.04 | 0.83 |
| Ar+4%$D_2$ | 1.07 | 0.84 |
| Ar+$H_2O$ | 0.94 | 0.76 |
| Ar+$D_2O$ | 1.00 | 0.81 |

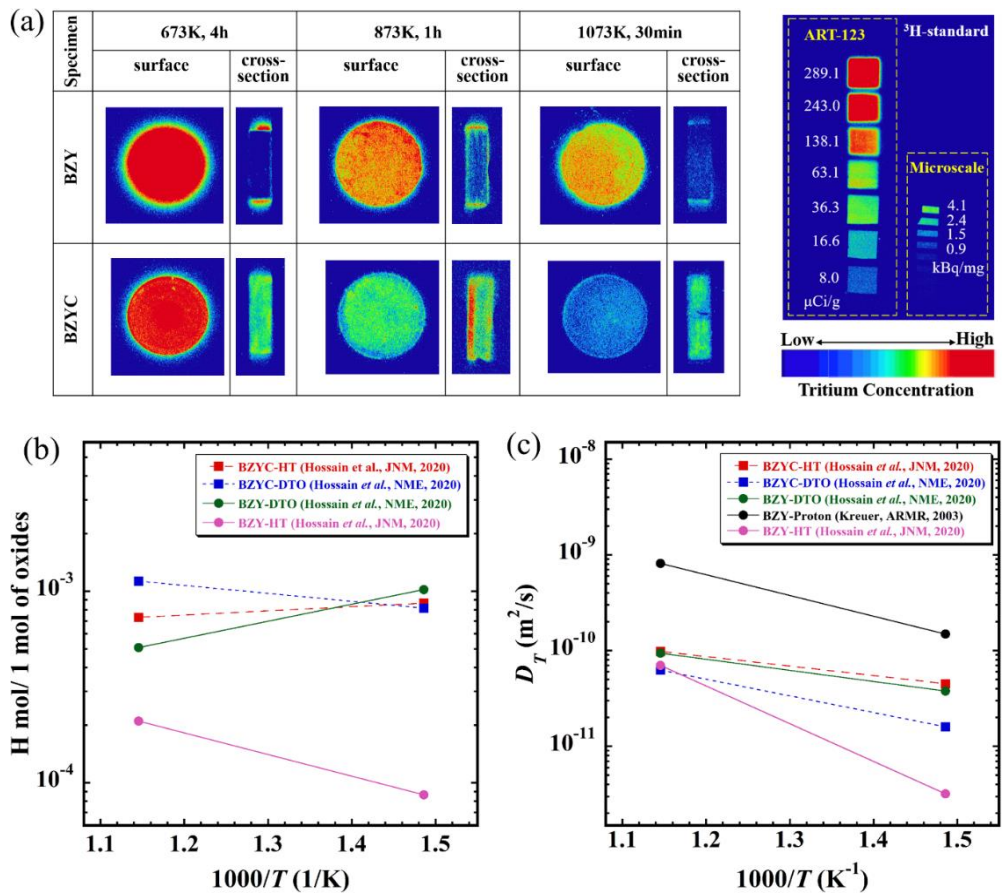

**Fig. 1**. (a) Hydrogen isotope (tritium) IP images of the cross-sections and outer surfaces of the BZY and BZYC materials represents the distribution of hydrogen, (b) Arrhenius plot of the hydrogen solubilities for the BZY, and BZYC, and (c) Arrhenius plot of the hydrogen diffusivities for the BZY, and BZYC [5,6].

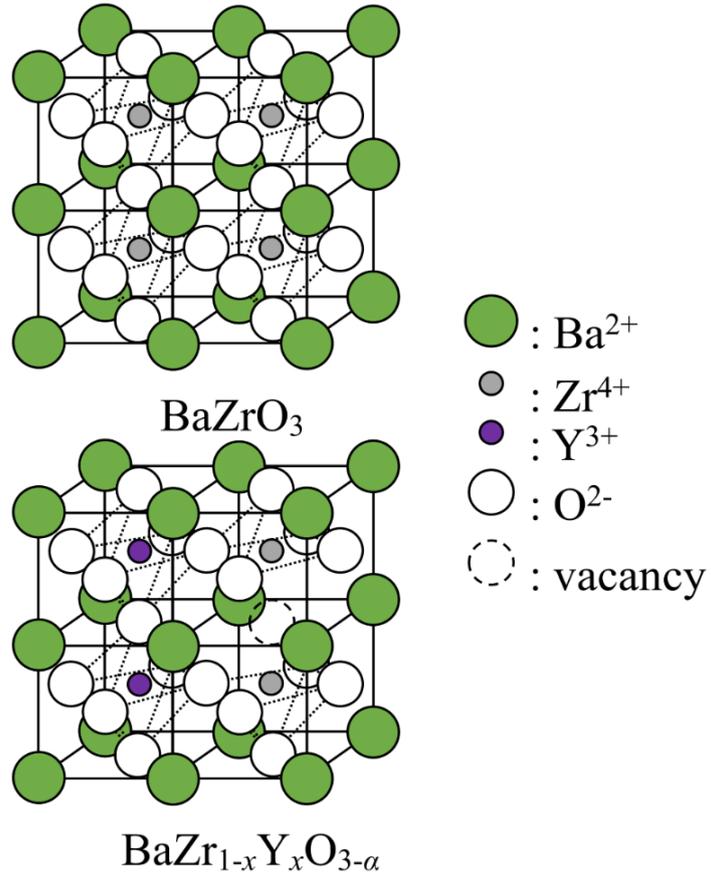

**Fig. 2**. Structure of $BaZr_{1-x}Y_xO_{3-\alpha}$.

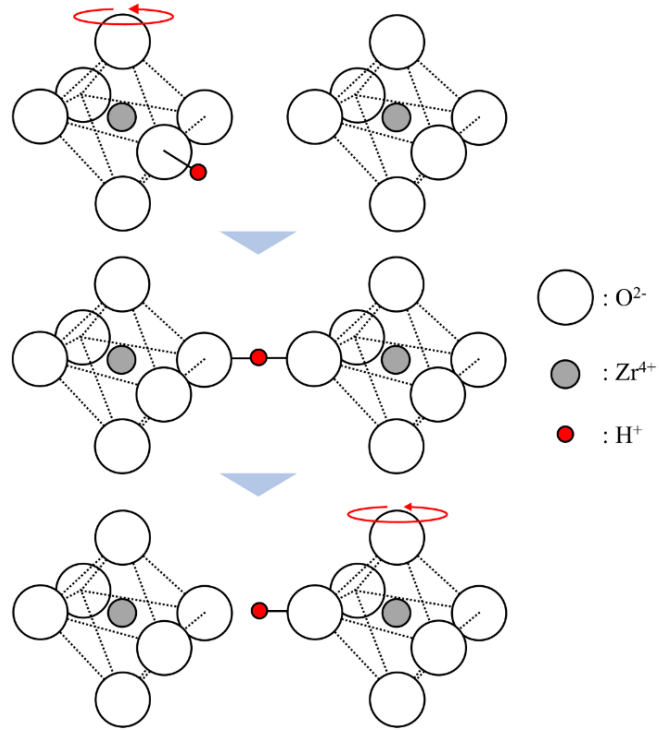

**Fig. 3**. Proton conduction by Grotthuss mechanism.

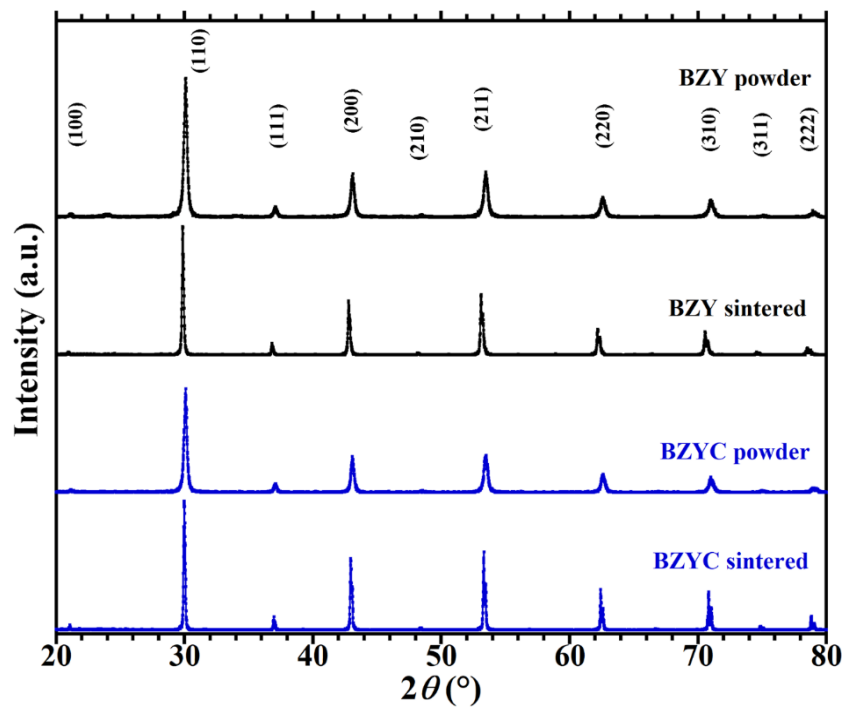

**Fig. 4.** XRD patterns for the powder and sintered samples of BZY, and BZYC.

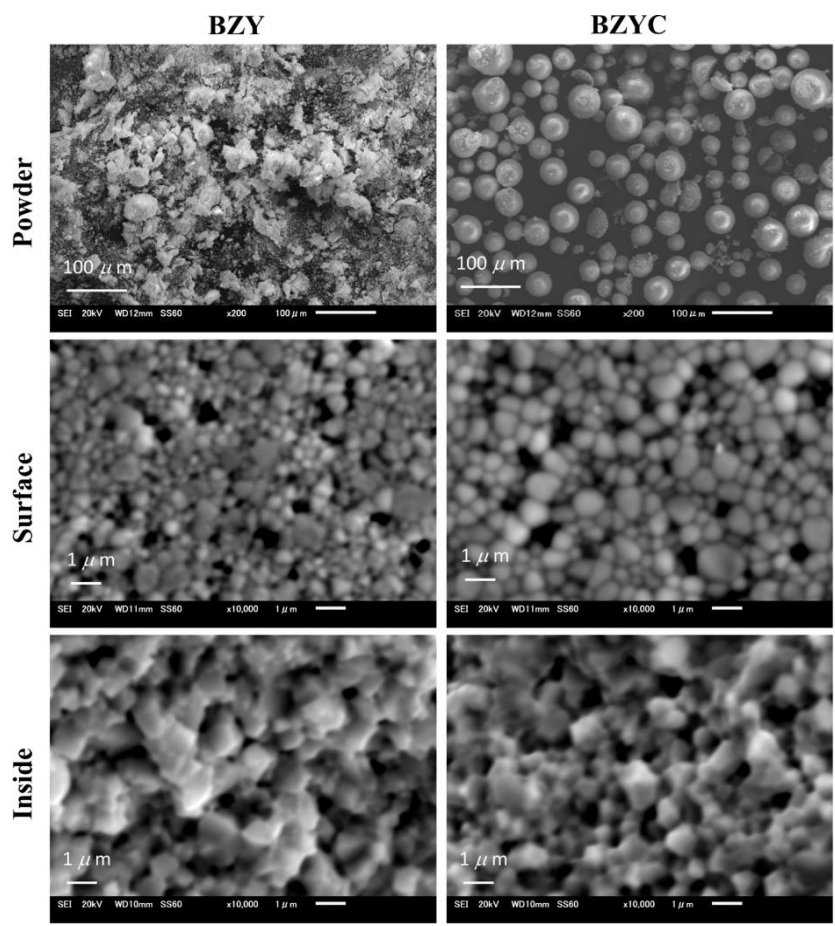

**Fig. 5.** SEM images for powder, sintered outer surface, and sintered inside (fractured surface) of BZY, and BZYC.

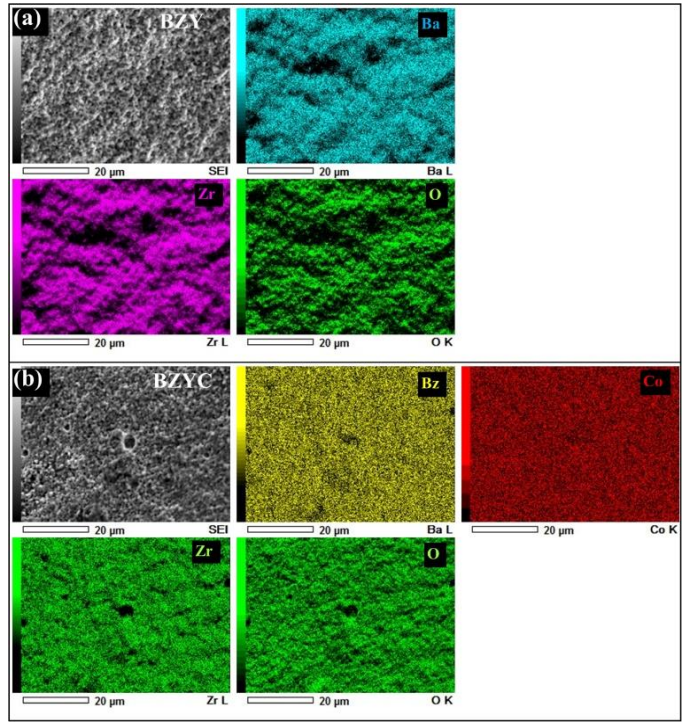

**Fig. 6**. EDX mapping for the inside of the BZY, and BZYC sintered samples.

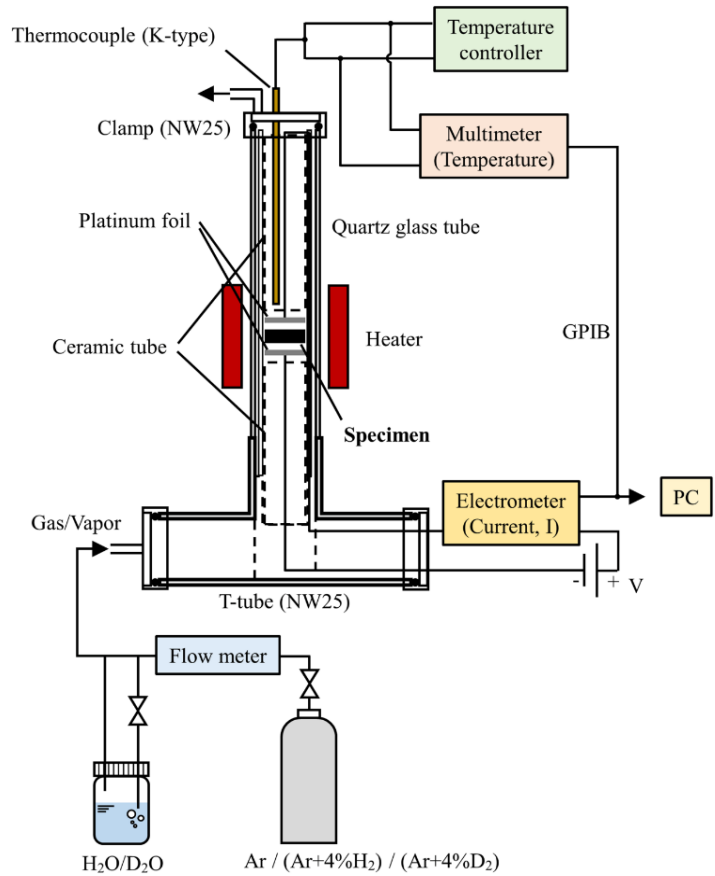

**Fig. 7.** Schematic diagram of the conductivity measurement device.

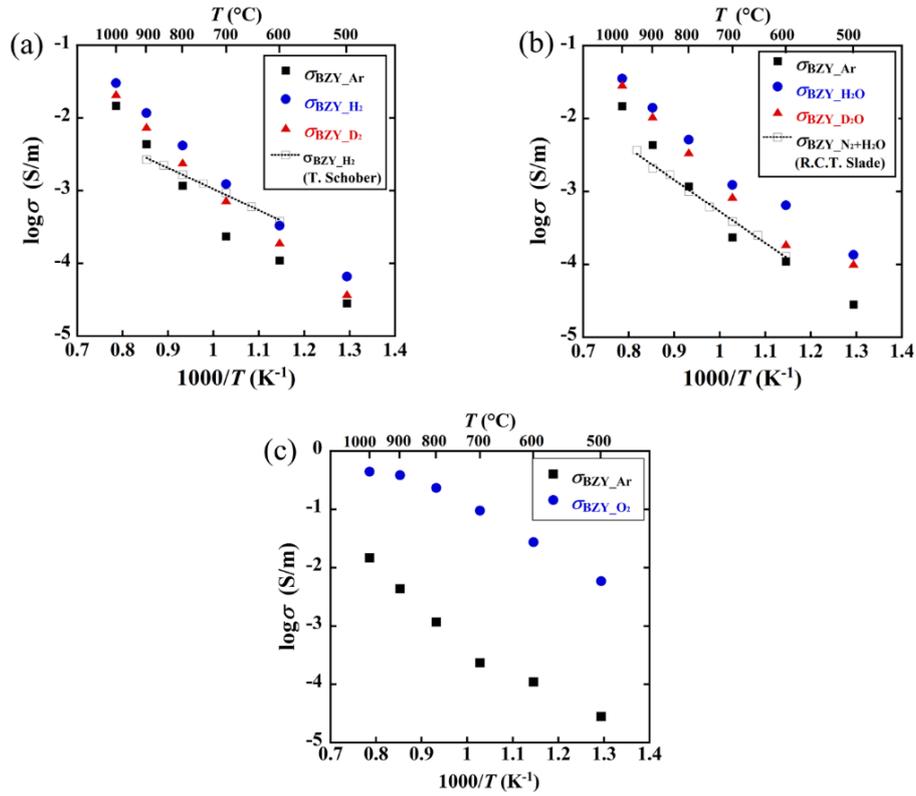

**Fig. 8**. Conductivity of BZY under: (a) Ar, (Ar + 4% $H_2$), and (Ar + 4% $D_2$) atmospheres, (b) Ar, (Ar + 4% $H_2O$), and (Ar + 4% $D_2O$) atmospheres, and (c) Ar and $O_2$ atmospheres.

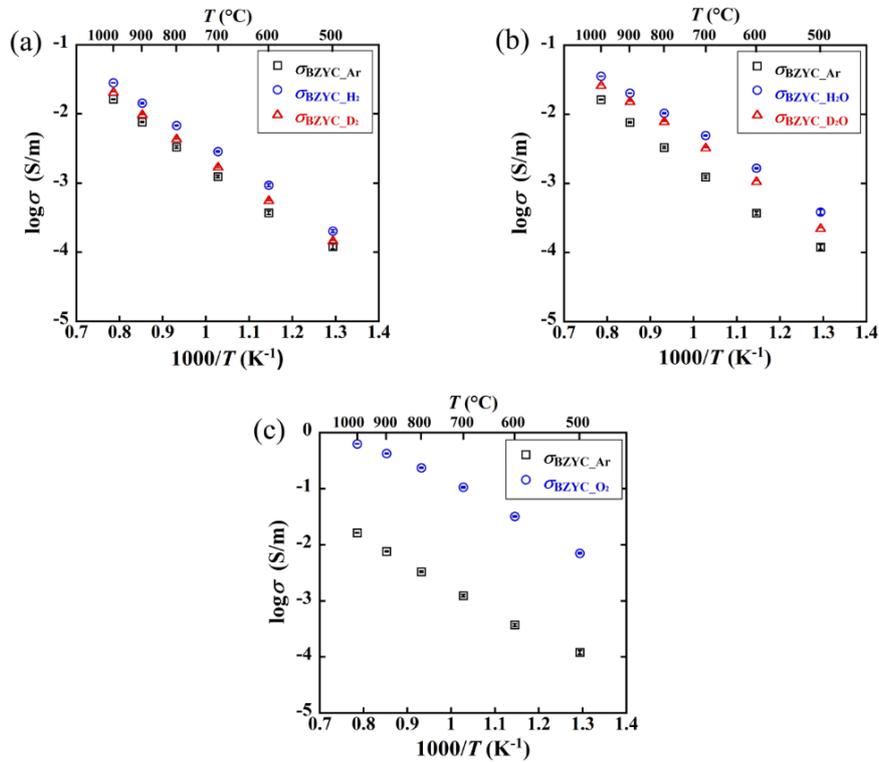

**Fig. 9**. Conductivity of BZYC under: (a) Ar, (Ar + 4% $H_2$), and (Ar + 4% $D_2$) atmospheres, (b) Ar, (Ar + 4% $H_2O$), and (Ar + 4% $D_2O$) atmospheres, and (c) Ar and $O_2$ atmospheres.

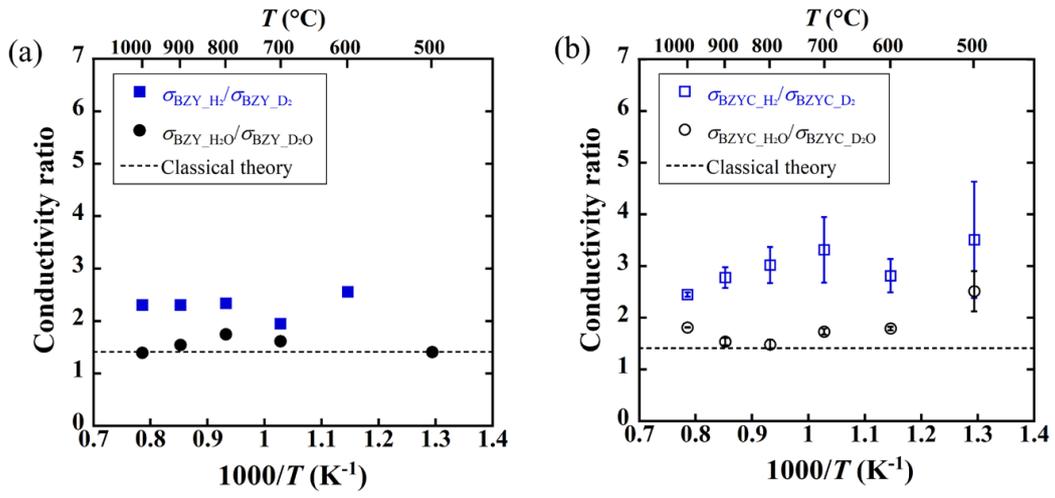

**Fig. 10.** Conductivity ratio of BZY and BZYC samples.